\documentstyle[12pt,a4,epsfig]{article}
\newcommand{\epem}{{\ifmmode e^+e^-  \else $e^+e^-$\fi}}
\newcommand{\pperp}{{\ifmmode p_t  \else $p_t$\fi}}
\newcommand{\sqrts}{{\ifmmode \sqrt{s} \else $\sqrt{s}$\fi}}
\newcommand{\zz}{{\ifmmode Z^0  \else $Z^0$\fi}}
\def\bbar{{\mathchar'26\mskip-9mub}}
\begin{document}
 
\begin{flushright}
MPI-Ph/94-76\\
November 1994\\
\end{flushright}

\bigskip

\bigskip 

\bigskip 

\bigskip 

\begin{center}
\large {\bf Directions in High Energy Physics}\\
\vspace{3cm}
\normalsize
{\bf Bodo Lampe}\\
 
\bigskip
Max-Planck-Institut f\"ur Physik, Werner-Heisenberg-Institut\\
F\"ohringer Ring 6, D-80805 M\"unchen\\
e-mail: bol@mppmu.mpg.de\\
\vspace{4cm}
\end{center}
\begin{center}
{\bf Abstract}
\end{center}
\bigskip
\noindent The future goals of particle physics are
classified from a theorist's point of view.
The prospects of mass and mixing angle determination
and of the top quark and Higgs boson discovery are discussed.
It is shown that the most important progress will come
from LHC and NLC. These machines should be planned and
developed as quickly as possible.
\newpage
\bigskip
{\bf 1. Introduction}
 
Physics is the science in which matter and ratio are most
tightly connected. High energy physics is its most fundamental
branch because we try to find ever smaller constituents
of matter. It was born from nuclear physics as higher energies
became available. Its goal has always been to discover
new particles and determine their properties in high energy
collisions. The properties to be determined are mass,
quantum numbers and couplings to other particles. When
it comes to still higher energies further questions can be examined,
like compositeness or more fundamental symmetries. Finally,
a complete theoretical model for the particle can be developed,
based in the best cases on simple and fundamental mathematical
principles.
 
I am sure that this concept of high energy physics will successfully
persist in the future although new experiments become increasingly
expensive. Most important items are the discovery and examination of the
top quark and of the Higgs particle because this will open new frontiers
in our understanding of the fundamental interactions. It would be
desastrous  if we would wait and withdraw to cheaper but less
important projects. Less important projects also cost money,
bind energy and, most of all, they distort the direction of research.
I can see a tendency for this in some
recent decisions and I want to warn that this could jeopardize
the future of particle physics. Once money is lost in wrong directions
it is difficult to attain new money, even for important experiments.
It is the main aim of this article to discuss these issues in detail
and to rate the various projects according to their priority.
 
A standard popular objection to basic science is that the number
of physical laws is limited and most of them are already known.
I cannot see that this is an argument against high energy physics.
As will be discussed in the next section it is absolutely clear
that the deepest physical laws have not yet been found. In fact,
the next section will start with a summary for and against the
so-called standard model of elementary particle physics.
Afterwards, a set of experiments as suggested by the standard
model will be valuated. We shall see that the heavy particles of the
standard model are at the centre of interest and should be treated with
the highest priority. Finally, I shall come to physics beyond
the standard model and to the question how reasonable
certain specific null
experiments are which search for deviations from the standard model.
Usually, the theoretical ideas on which such experiments are
based are rather exciting but the performance is boring when nothing
is found.
 
Note that this paper is not intended as an exhaustive review
but as a grading of present and future high energy physics
experiments. Its aim is to initiate discussions about the future
directions and to sharpen the view for what is important and what not.
 
The known elementary particles do not form a very complex system.
As compared to biological systems, for example, they are remarkably
simple. Complex systems are always multiparticle character whereas
high energy physics is eventually searching for the
most fundamental entities and their interactions and is thus
different from other branches of science like chemistry,
astrophysics or even mathematics. This should be kept in mind when
the significance of experimental and theoretical ideas is reflected.\\
 
{\bf 2. The standard model}
 
\begin{figure}
\begin{center}
\epsfig{file=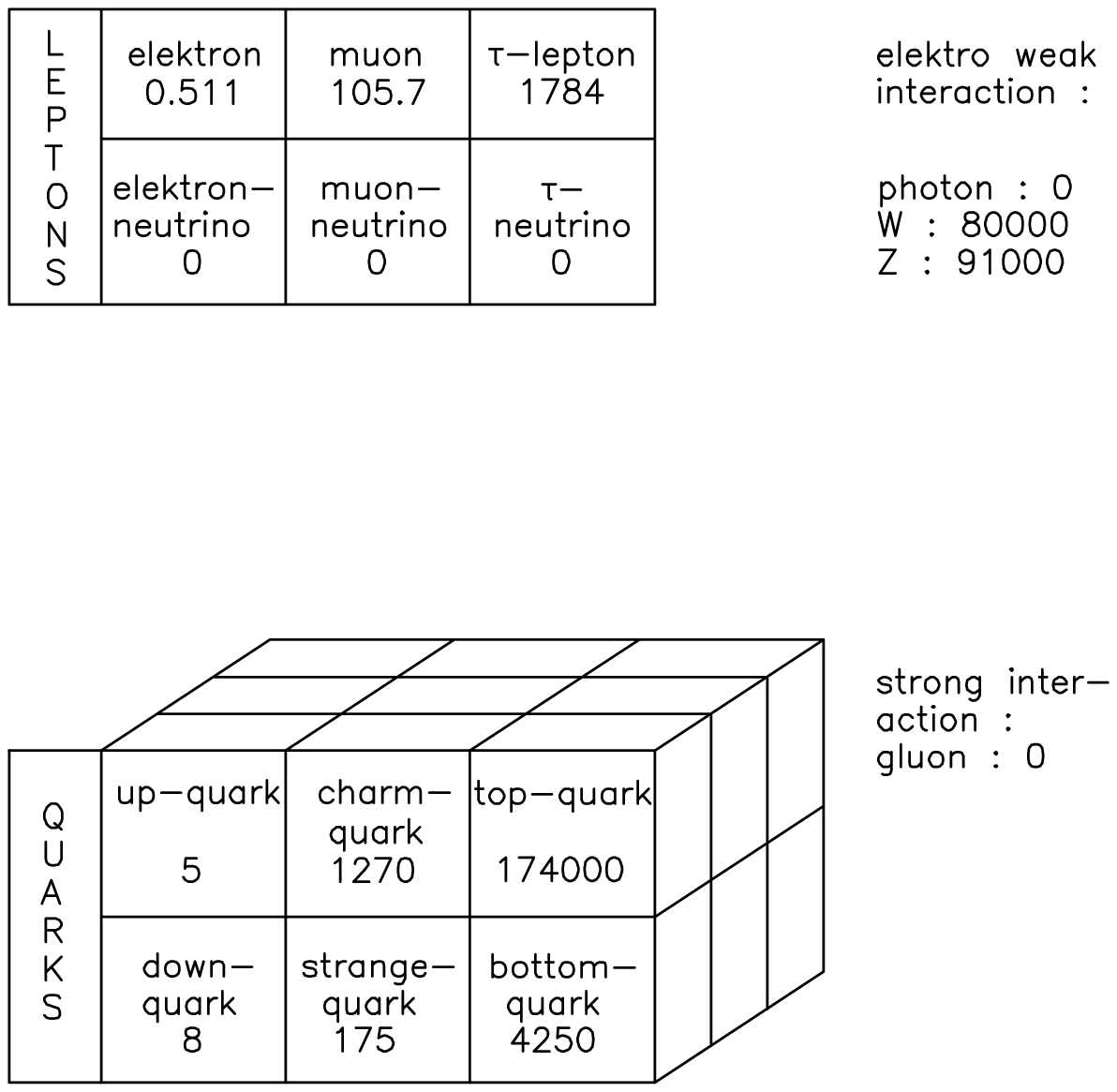,height=8.5cm}
\vskip 0.5cm
\caption{}
\label{fig0}
\end{center}
\end{figure}

The standard model of elementary particle physics describes the
interactions of the quarks and leptons as mediated by the vector
bosons of the strong (gluons), weak ($W^\pm,Z$) and
electromagnetic (photon) interactions, c.f.\ fig. \ref{fig0}.
Quarks and leptons are fermions (= spin$\frac{1}{2}$ particles)
whereas the vector bosons have spin~1. Finally, there is the
Higgs boson, a spin~0 particle, which is associated with
the generation of all particle masses. It is certainly the most
mysterious object of the standard model
because it is as yet undiscovered
and relies on a purely theoretical construction about symmetry
breaking. Once discovered, the nature of the Higgs particle will be
established by demonstrating that its couplings to other particles
grow with their masses.
 
The standard model has been very successful phenomenologically.
I know of no
experiment which definitely contradicts the standard model.
However, it is not believed to be
the ultimate theory of nature, because it has too many
free parameters, namely, all the fermion masses.
The standard model describes particles
whose masses range between 0 (neutrinos and photon) and
100-1000~GeV
and one would like to understand better, how these vastly different
masses arise and whether there are other important mass scales
at higher energies. Some related questions are:
the masslessness of the neutrino and the stability of the proton.
These items will be discussed in later sections.
 
The gauge sector of the standard model is remarkably clear
and extremely restricted, with only 3 dimensionless
coupling constants describing the interactions of the vector bosons with
themselves and with the fermions. Through this
one was able to predict many
properties of $c$-, $b$- and $t$-quark and the $\tau$-lepton long
before these particles were actually discovered.
It is even possible that the standard model gauge group is
embedded in a larger more simply connected group. This is the
``Grand Unification'' scenario first suggested by
Georgi and Glashow.\\
 
{\bf 3. Vectorboson self couplings from LEP1 and LEP2}
 
\begin{figure}
\begin{center}
\epsfig{file=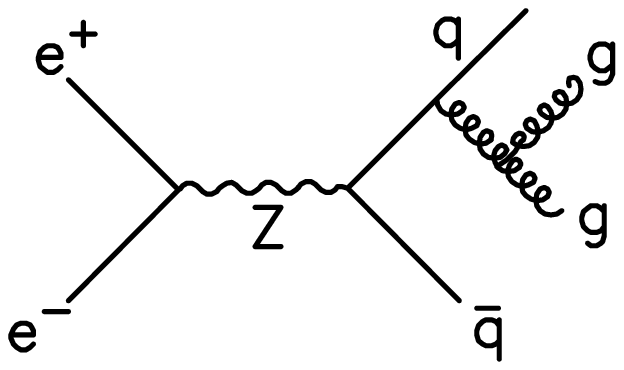,height=3cm}
\vskip 0.5cm
\caption{}
\label{fig1}
\end{center}
\end{figure}

The direct self coupling of vector bosons is a firm and
pronounced prediction of nonabelian gauge theories. It
makes the behavior of the gluon, the $W$ and the $Z$ much different
from that of the photon. For example, a large amount of the 4-jet
cross section in $\epem$ annihilation should come from processes
in which one gluon splits into two (fig. \ref{fig1}).
This effect can quantitatively
be tested because the LEP1 experiment has reached a rather high
precision concerning its 4-jet cross section. On the theoretical
side the calculation of higher order corrections to
$d\sigma_{{\rm 4jet}}$
is still missing. This is unfortunate because this quantity
offers the cleanest possibility to check the triple gluon coupling.
Information can also be obtained from proton collisions -- but
with much higher uncertainty (due to our ignorance of the gluon
distribution at small $x$).
 
\begin{figure}
\begin{center}
\epsfig{file=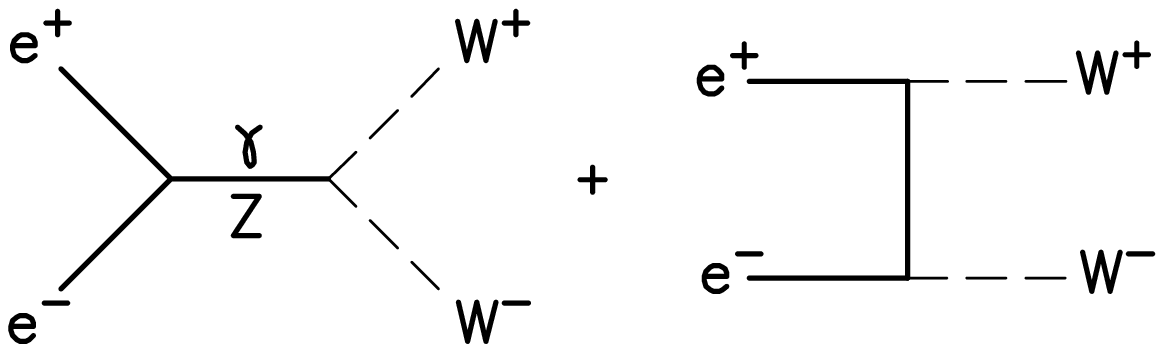,height=3cm}
\vskip 0.5cm
\caption{}  
\label{fig2}
\end{center}
\end{figure}

In spite of the high precision of the LEP1 experiment the 3-gluon
self coupling will always remain a little in the dark. This is
because of the usual problem of QCD, the washing out by
hadronization effects. In this respect the situation is better for
the $ZW^+W^-$ interaction which will be studied at LEP2 [1] 
via the
process $\epem\to W^+W^-$ (fig. \ref{fig2}).
I consider this experiment very important.
It is a fundamental experiment, in the sense that
one can hope for deviations from the standard model, and it must be
made by all means to fix once and for all any doubts which are left
concerning the description of the standard model gauge sector.
It can be shown that the sensitivity of the experiments to the
$ZW^+W^-$ coupling increases with energy so that an $\epem$
machine at total energy 500~GeV would be really superior
to LEP2 ($\sim$ 200~GeV). This is a first strong argument for an
$\epem$ collider at ultrahigh energies whose prospects will be
discussed in section~7.\\ 
 
{\bf 4. Determination of parton densities at HERA}
 
Proton collisions play an eminant role for the discovery of
new particles. However, cross section predictions for proton
colliders are plagued by a number of uncertainties, in particular
concerning our ignorance of the internal structure of the
proton. Deep inelastic lepton nucleon scattering can be reliably
used to determine the structure of the proton in terms of its
quark and gluon constitutents, i.e.\ the parton distribution
functions. If these functions are determined accurately,
one can obtain quantitative predictions for all sorts of cross
sections in proton proton collisions, like top quark or
Higgs boson production or the production of exotic particles.
At LHC with its high beam energy the behavior of the parton
distributions at small $x$ becomes essential because partons
with a relatively small amount $x$ of the proton momentum can produce
the new particles. The dominant parton distribution at small $x$
is the gluon distribution. The HERA experiment at DESY is able to
measure in the small $x$ region ($x\sim 10^{-3}$ to $10^{-2}$)
whereas
pre-HERA data on deep inelastic scattering could not resolve the
gluon density function at small $x$. In this sense HERA is a very
reasonable experiment whose result will be used until the next century.
Unfortunately, the theoretical handling of the very small $x$ ($<10^{-3}$)
HERA data is not fully clear. At very small $x$, there are nonperturbative
gluonic effects besides the perturbative ones. A complete theoretical
picture of how to combine these effects is still missing.
 
It should be stressed that the determination of parton
densities is the only important fundamental contribution which can
be expected from HERA. Certainly, a lot of interesting physics like
photoproduction, jets or
heavy quarks can be done and I don't
want to lessen their merits. But the real justification for the
large and expensive HERA project comes from the structure function
determination. It is very important that this is done as
precisely as possible.

It should also be stressed that HERA is completely
a standard model machine.
It is unlikely that nonstandard physics will be found at this
experiment.\\
 
{\bf 5. Light fermion masses and CKM matrix elements}
 
The masses of the standard model
particles can be deduced from experimental observations and at present
cannot be explained theoretically. Due to QCD effects, the quark
masses are much less accurately known
than the lepton masses. In fact, there is still a discussion going
on about the values of the light quark masses $m_u,m_d$ and $m_s$, and
it is questionable whether they can ever be determined to better that
20 \%.
 
An additional complication arises in the quark sector because the
various flavors mix, i.e.\ the physical ``mass'' eigenstates
are different from the interaction eigenstates.
As a consequence, a nondiagonal $3 \times 3$ matrix $V$ appears in the
charged current. $V$ is called the CKM matrix and it can be shown
theoretically that it has four independent parameters of physical
significance. A lot of physicists
in various experiments work on their experimental determination.
It is reasonable to say that rather precise values for
all four parameters will be available around the 
beginning of the next century.
Very precise information will come from the two ``$b$-factories''
[2] 
built in America and Japan which use the advantages of the process
$\epem\to b\bbar$ to determine $V_{ub}$, $V_{cb}, m_b,m_c,\ldots$.
This is an important issue because the masses and
mixing angles are the largest subset
of free parameters in the
standard model and
should be determined very precisely. It will be the basis for
future physical analyses, theoretically and experimentally, within
and beyond the standard model.\\
 
However, I do not understand why two (or even three?) $b$-factories
are built where one would be sufficient.
The political explanation for this wrong decision is that
in the various countries one is afraid to approve very large projects
(like the NLC). I am sure that in the long run this will prove
harmful because it misdirects tight ressources and could even jeopardize
the future of high energy physics.\\ 
 
{\bf 6. LHC, the top quark and the Higgs boson}
 
After the cancellation of the SSC the LHC (``Large Hadron Collider'')
at CERN [3] 
is the only remaining high energy proton collider project.
It will operate at an energy of about 16~TeV and will allow to
discover the Higgs particle as well as solidly establish the
existence of the top quark. The machine also provides unique
opportunities to search for new heavy particles as predicted by
theories beyond the standard model. The LHC will probably  
become one of the most rewarding project in the history of high energy
physics. 
 
The $\bar p p$ colliders which have operated in the last decade
at CERN and Fermilab have been very successful in discovering heavy
fundamental particles (the $W$ and $Z$ boson and the top quark).
The LHC with its much higher energy will continue
this success because it allows to produce a large number of top
quarks and even Higgs bosons so that the decay properties
of these particles can nicely be studied. Furthermore, as
compared to high energy $\epem$ machines it has a better
capacity to produce and discover very heavy particles with masses
of order 1~TeV. Heavy particles, within and beyond the standard
model, offer the best possibility to open really new frontiers
in physics, because
they have never been produced on earth and some suprise is likely
to arise when they are discovered and examined. Therefore the
LHC project is of extreme importance for the future of physics.

It is not so bad that the SSC plans have been cancelled because
the physics potential of LHC and SSC is rather similar. Furthermore,
LHC costs are much smaller than SSC, because the experiment
will take place in an existing tunnel (the LEP tunnel).
 
\begin{figure}
\begin{center}
\epsfig{file=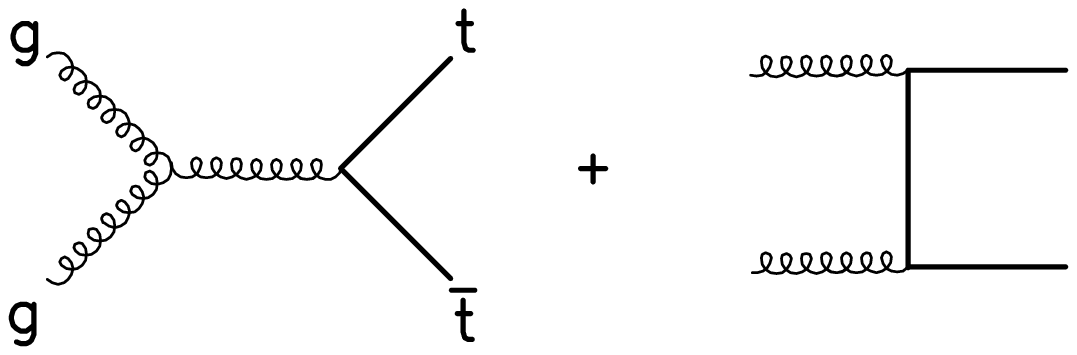,height=3cm}
\vskip 0.5cm
\caption{}
\label{fig3}
\end{center}
\end{figure}

\begin{figure}
\begin{center}
\epsfig{file=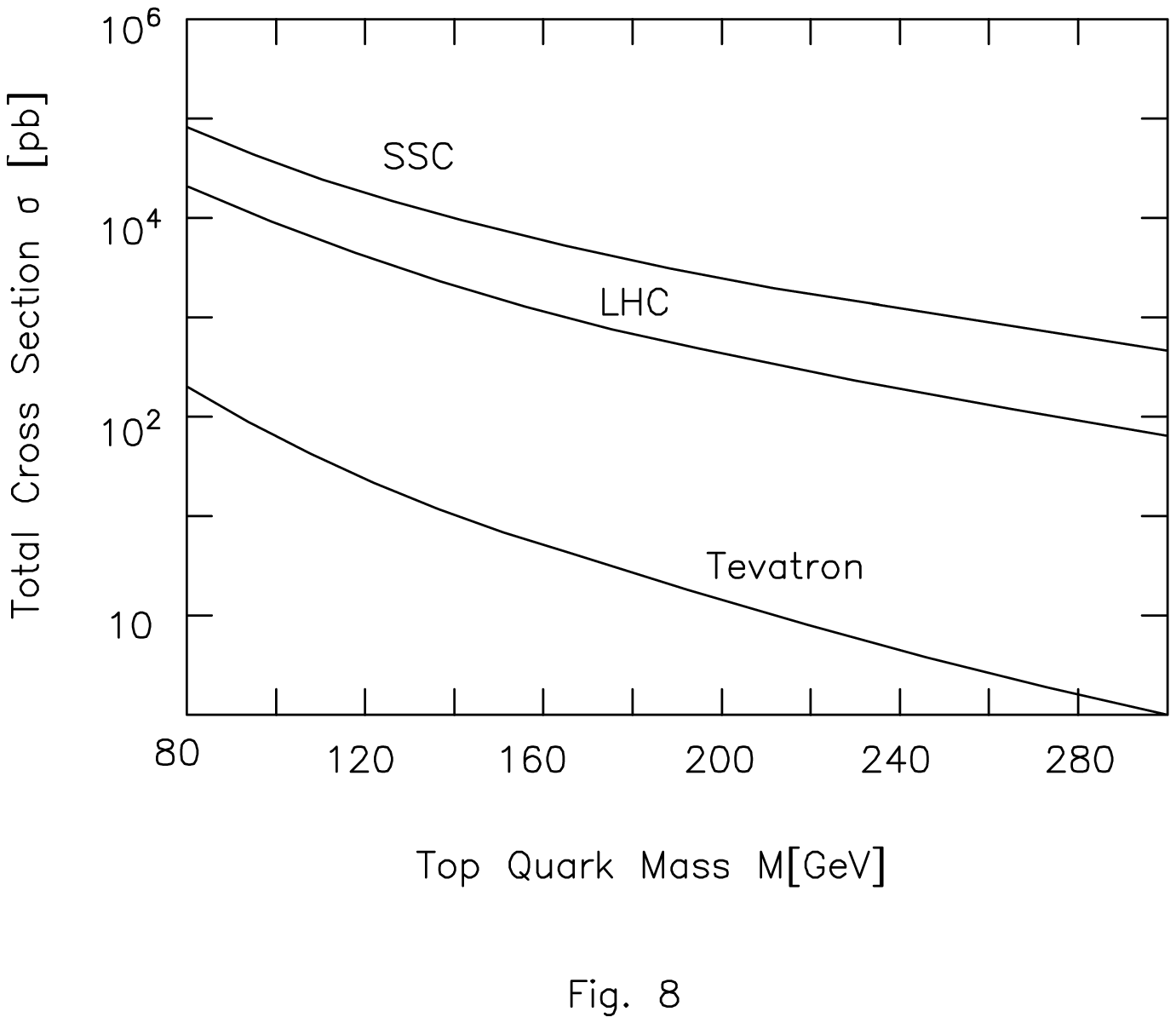,height=8cm}
\vskip 0.5cm
\caption{}
\label{fig4}
\end{center}
\end{figure}

The main production mechanism of top quarks at LHC is gluon-gluon fusion
$gg\to t\bar t$ (fig. \ref{fig3}).
Estimates of cross sections are shown in fig. \ref{fig4}.
They correspond to a number of about 100000 top quarks to be
created per year
(for a top mass of 175~GeV). The main uncertainty in these cross
section estimates arises from ignorance about the parton
distributions.
 
\begin{figure}
\begin{center}
\epsfig{file=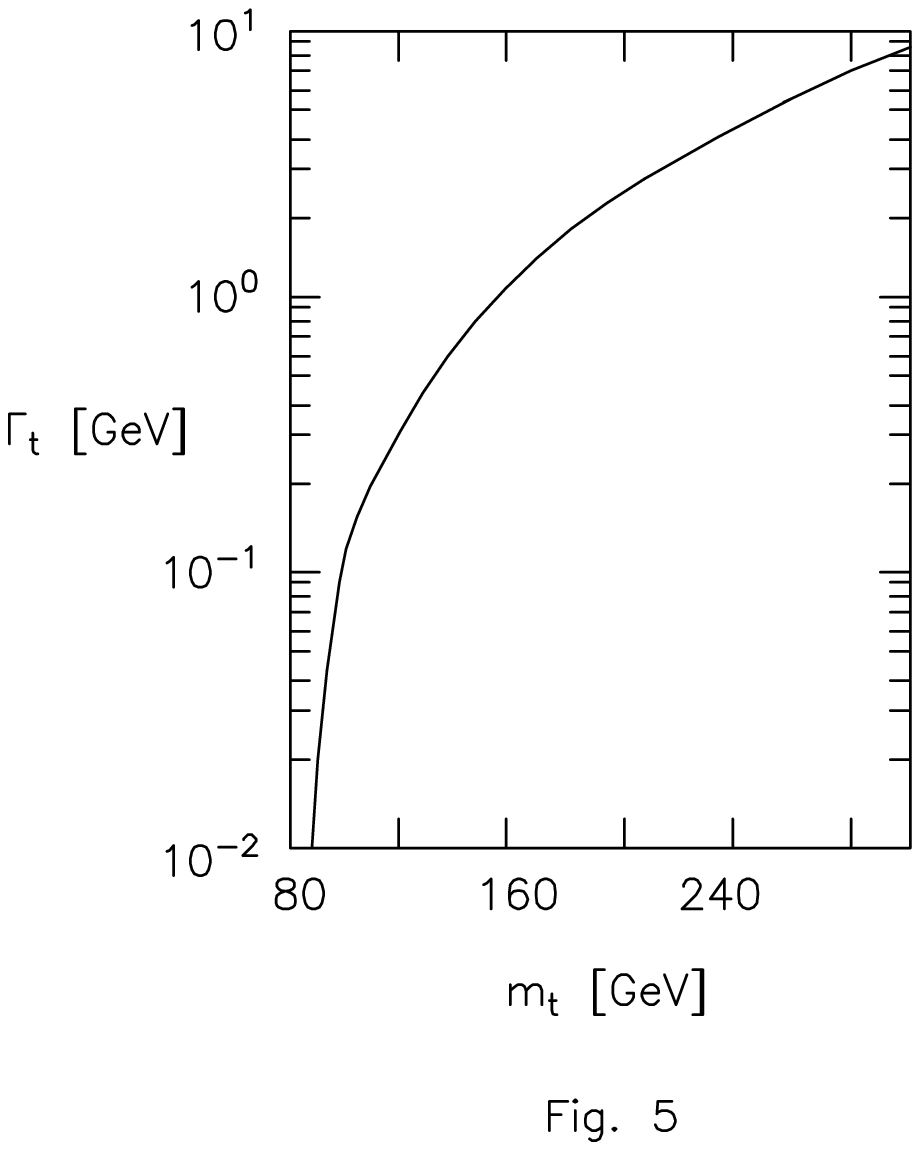,height=7cm}
\vskip 0.5cm
\caption{}
\label{fig5}
\end{center}
\end{figure}

The main decay mode ($>$ 99.8\%) of the top quark is
$t\to b+W^+$ and it leads to a width of about 1.5~GeV (see fig. \ref{fig5}).
The semileptonic branching ratio is
$BR(t\to\mu^+)={1\over 9}$
leading to the emission of clearly visible hard isolated leptons.
Additional leptons from semileptonic $b$-quark decays are softer
and non-isolated, i.e.\ associated with hadronic jets.
 
Within the standard model,
the only parameter for the top quark to be fixed is the mass.
It can be determined by studying the invariant mass distribution of
the jets recoiling against the lepton in one-lepton events.
This way a top quark mass value of about $(175\pm 20)$~GeV
has been determined at the Tevatron and an error of 1~GeV
or less can be envisaged at LHC.
Beyond the standard model there is a variety of possibilities, like
deviations from the V-A coupling, effects of charged
Higgs bosons in top decay, supersymmetry etc. which can all
be searched for.
 
\begin{figure}
\begin{center}
\epsfig{file=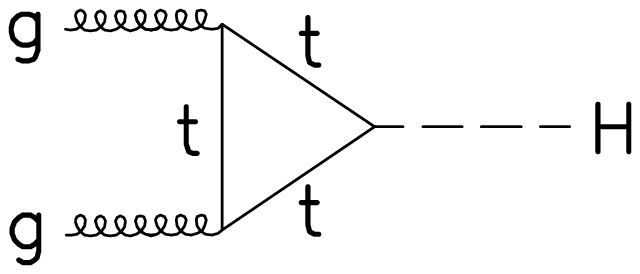,height=3cm}
\vskip 0.5cm
\caption{}
\label{fig6}
\end{center}
\end{figure}

\begin{figure}
\begin{center}
\epsfig{file=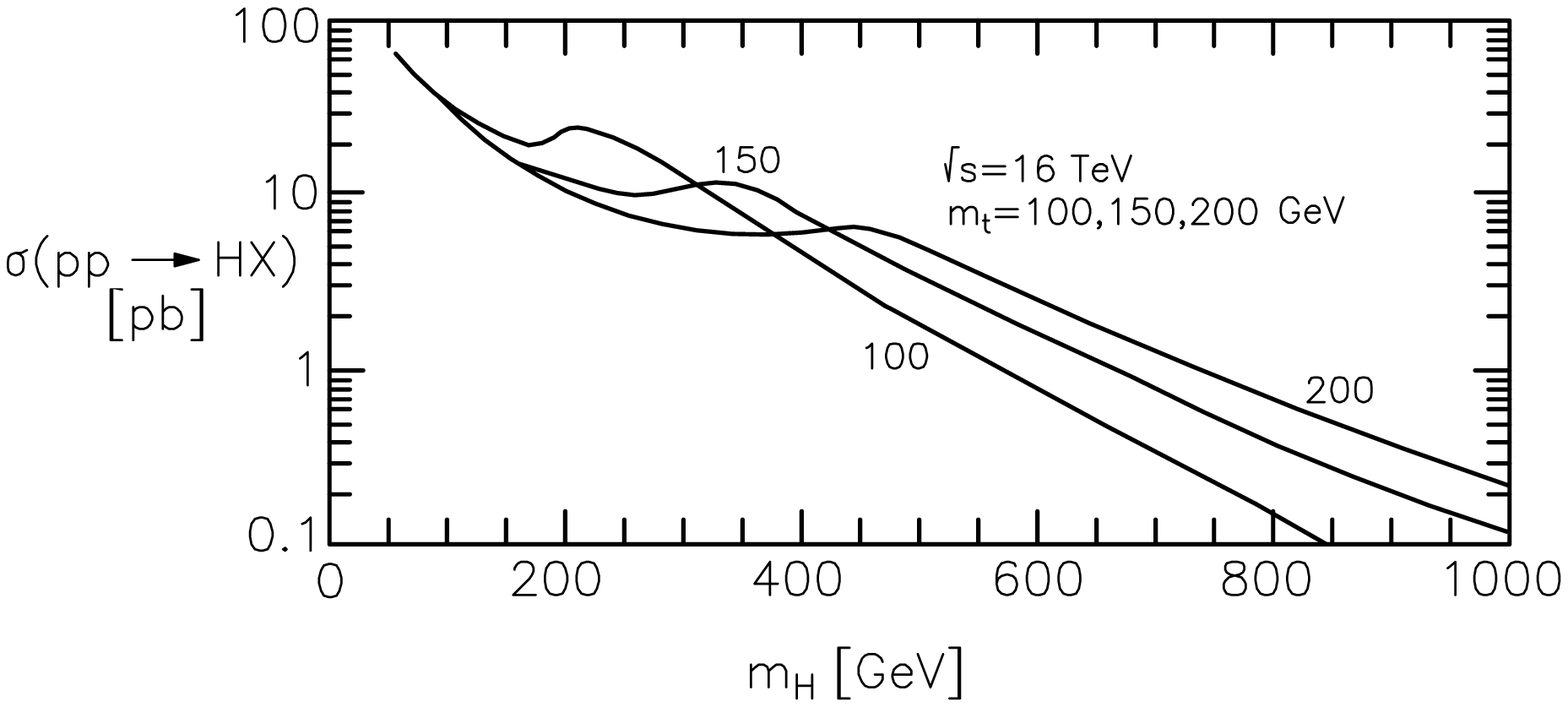,height=6cm}
\vskip 0.5cm
\caption{}
\label{fig7}
\end{center}
\end{figure}

The main production mechanism of the Higgs boson at LHC
is the fusion mechanism $gg\to H$ with an
intermediate top quark triangle (fig. \ref{fig6}).
Estimates of the cross section
are shown in fig. \ref{fig7}.
We see that the cross section is typically a factor of 100 smaller
than for the top quark, but still a lot of Higgs particles
should be produced at LHC.
 
\begin{figure}
\begin{center}
\epsfig{file=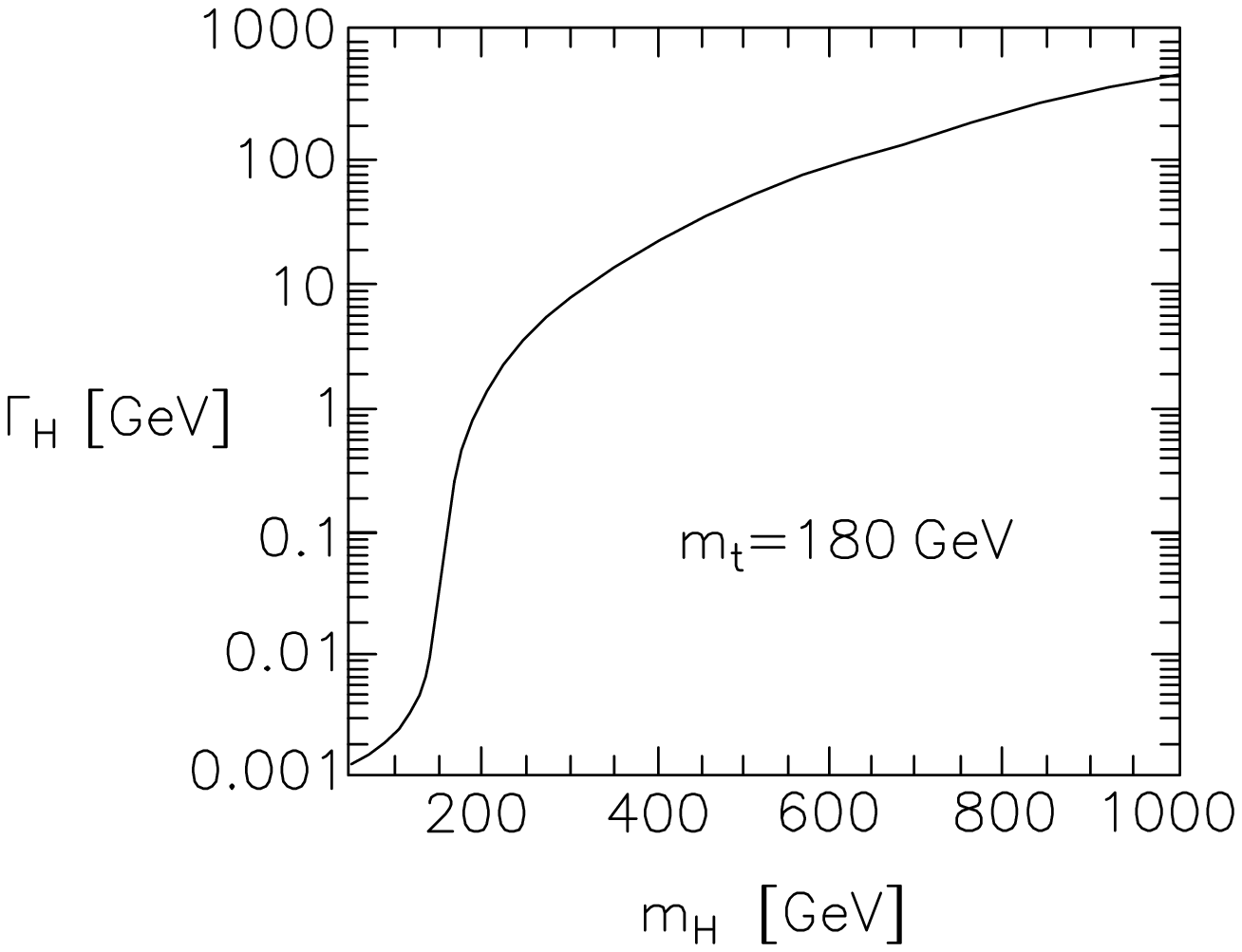,height=7cm}
\vskip 0.5cm
\caption{}
\label{fig8}
\end{center}
\end{figure}

The main decay modes of the Higgs boson are $H\to b\bar b$ resp.\
$WW/ZZ$ depending whether $m_H <2m_W$ or $>2m_W$ and they lead to a
width shown in fig. \ref{fig8}.
Due to large backgrounds from ordinary $b\bbar$ or $WW/ZZ$ events
the search for the Higgs boson at LHC is difficult. -- But it is not
hopeless because it should be possible to see the Higgs events as
bumps on invariant mass distributions for those processes.
A Higgs boson with mass of 0~(100~GeV) seems to be particularly
difficult to detect because of the large $b\bar b$ background.
I am optimistic that even that case is tractable because proton
collisions are usually more successful than anticipated.
 
Within the standard model
the only parameter for the Higgs search experiment
to be fixed is the Higgs mass.
$m_H$ is unknown but should be well
below 1~TeV, for theoretical reasons. Furthermore,
LEP1 results restrict $m_H$ to $m_H>65$~GeV.
At LHC, $m_H$ can be determined from the location
of the bump in those invariant mass distributions.

Together with
the Fermi constant the Higgs mass
completely fixes the form of the Higgs
potential in the standard model.
It is very important to know $m_H$ and to check the high
energy tail of the standard model. Of course,
more than one Higgs field and correspondingly more
free parameters in the Higgs potential would complicate the
situation. In this case the NLC (next linear $\epem$ collider)
would be extremely useful for clarification.\\ 
 
{\bf 7. NLC, t and H}
 
The next linear $\epem$ collider (NLC) [4] 
expected to operate in
the energy range between 300 and 800 GeV will allow to perform very
precise studies of the heavy particles in the standard model,
the top quark, the electroweak bosons and the Higgs particle.
The machine will also provide unique opportunity for new
physics searches.
 
The $\epem$ colliders which have operated over the past two decades
have been outstandingly successful in exploring the fundamental
interactions and constituents of matter. The charm quark, the
$\tau$ lepton, the gluon and the bottom quark were discovered
and established by SPEAR, PEP1 and PETRA, and their properties could
be studied in a clean experimental environment. Later on,
the precision
analysis of the $Z$-boson and its decay modes at LEP has
established the validity of the standard model to a very high level of
accuracy.
 
The next generation of $\epem$ colliders will undoubtedly
continue this success story and reveal much about the properties
of the Higgs boson and the top quark.
As compared to a high energy proton collider the
production of these particles in $\epem$ annihilation can be studied
at a much higher level of precision. This way the standard model
predictions can be nicely checked and possible deviations from the
standard model
could be established. I have in mind here the pointlike
$V-A$
couplings of the top quark and the masslike couplings of the
Higgs boson, and, more in general, a precise examination of
the top Higgs connection, symmetry breaking mechanism etc.
 
In my opinion the NLC is an absolutely necessary project
to complement the LHC. No time should be wasted to
start on.
As time goes by, with smaller projects being approved, we might
be tempted to concentrate on them and withdraw from NLC and its
important questions. This would be desastrous because it would
jeopardize the future of particle physics.
 
\begin{figure}
\begin{center}
\epsfig{file=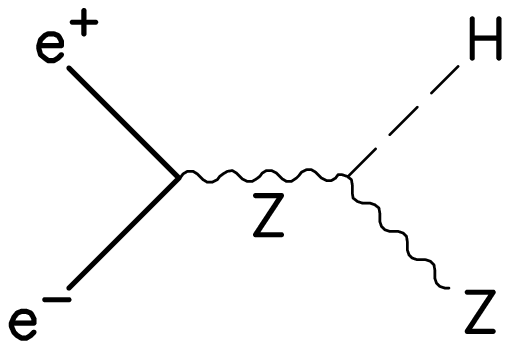,height=3cm}
\vskip 0.5cm
\caption{}
\label{fig9}
\end{center}
\end{figure}

\begin{figure}
\begin{center}
\epsfig{file=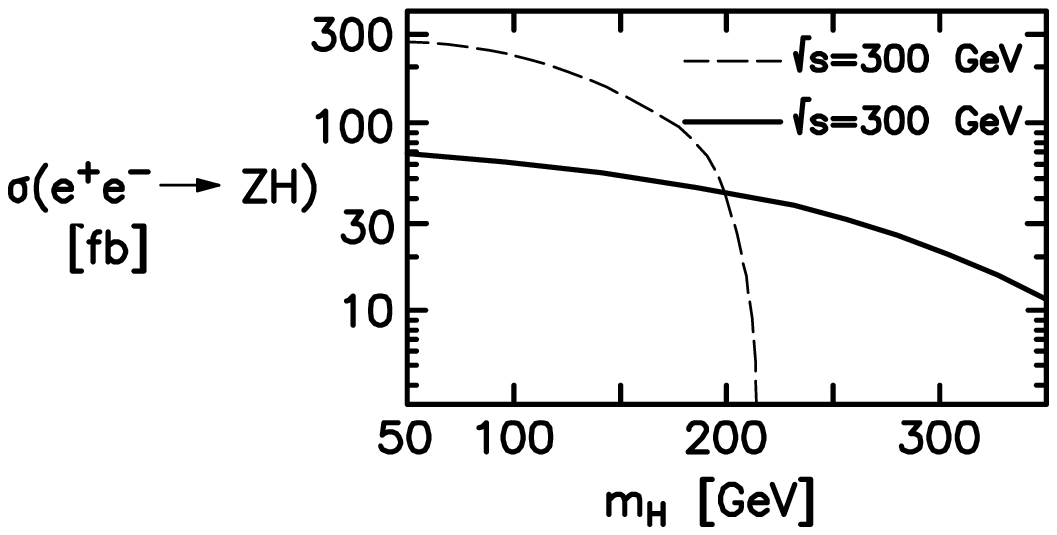,height=6cm}
\vskip 0.5cm
\caption{}
\label{fig10}
\end{center}
\end{figure}

The Higgs boson has not been found at LEP1. From this fact a
lower limit $m_H>65$~GeV can be deduced.
At LEP2 and NLC the $m_H$ range up to 90~GeV, resp.\ O(200~GeV) 
will be covered using the Higgs-strahlung process $\epem\to ZH$
(fig. \ref{fig9}).
This can be deduced from the production cross section fig. \ref{fig10}. 
NLC will be an ideal laboratory for the discovery of an ``intermediate''
O(100~GeV) mass Higgs field because it would
show up as a spectacular peak in the distribution of the
recoil mass of the Higgs-strahlung process. After the discovery
the properties of the Higgs boson can be accurately determined and
many
informations not available at LHC will be obtained. This will be the
basis for further tests of the standard model Higgs sector.
 
\begin{figure}
\begin{center}
\epsfig{file=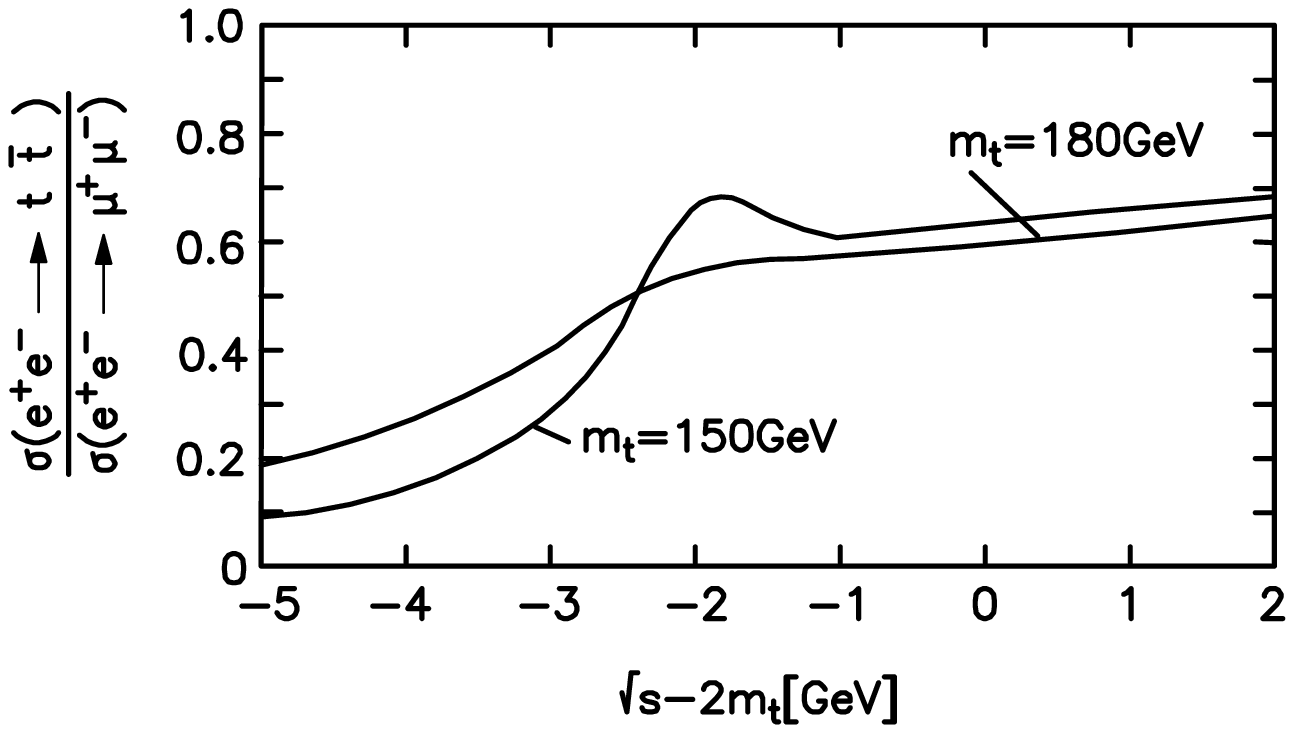,height=6cm}
\vskip 0.5cm
\caption{}
\label{fig11}
\end{center}
\end{figure}

Top Physics is also very interesting at the NLC. From the
behavior of the integrated cross section $\sigma (\epem\to t\bar t)$
in the threshold region one will be able to obtain an
extremely accurate value of the top quark mass (see fig. \ref{fig11}).
Now, $m_t$ is not a prediction of the standard model
but the couplings of the
top quark to the vector bosons $(W,Z,\gamma)$ are.
These couplings can be determined very precisely from differential
distributions of the $t\bar t$ decay products so that a high level
check of the standard model will be possible.
 
In summary, I would say that the NLC is the most interesting
high energy physics
project for the next century. The problems related to the NLC
are not on the physical but on the technological and financial side.
It is not fully clear whether $\epem$ beam energies of 500~GeV can
technically be reached but it will be
worth-while to overcome these difficulties because
the physics prospects are really fantastic.\\
 
{\bf 8. Physics beyond the standard model}
 
Theoretical developments beyond the standard model
are usually in a rather weak
position because of the lack of any experimental indication.
The most common are extensions of the standard model,
in the sense that these
theories have a slightly
enlarged symmetry or particle spectrum.
For example, grand unified theories (GUT)
are based on gauge groups like $SU_5$ which contain the
standard model
gauge groups $SU_3\times SU_2\times U_1$ as subgroup.
Consequently, the three running coupling constants $g_3, g_2$ and $g_1$
should converge to a common value $g_5$ at some large
``grand unification'' scale $M_X\approx 10^{14}$~GeV. The
measured values of $g_3, g_2$ and $g_1$ seem to indicate that this
may happen only if there is additional new
physics at scales
$\Lambda \ll M_X$, let's say $\Lambda\sim$~O(1~TeV) which would
modify the evolution of $g_{3,2,1}$
in the right way corresponding to somewhat larger scales
$M_X\approx 10^{16}$~GeV. Many theorists believe that this is
the so-called supersymmetry which would show up in the form that to
each existing particle a ``supersymmetric'' partner with
spin shifted by $\pm\frac{1}{2}$ and mass of order 1~TeV
exists. Personally, I see no compelling reason to believe
specifically in the supersymmetric scenario.
Nevertheless, it is an important issue to go to energies in
the TeV range and try to find new particles there.
Unfortunately, the GUT ideas do not give conclusive
evidence whether the new physics scale is 1 TeV, 10 TeV or
even 1000 TeV.It 
is not at all clear whether experiments at energies larger
than 100~TeV will ever be made. This is a question of the far future
which I cannot answer. They should be made, if they are technically
and financially possible.
 
A specific low energy prediction of GUT's is proton decay. It arises
through the existence of new superheavy $SU_5$ gauge bosons, of
mass $M_X$. The exchange of virtual $SU_5$ gauge bosons induces
baryon number violating processes, like $p\to e^+\pi^o$.
As compared to normal weak boson exchange processes they are suppressed
by a factor $(M_W/M_X)^4$. 

Proton decay experiments have been made
in the last two decades with increasing effort and without success.
From the present limit on the proton lifetime, $\tau_P\geq 10^{34}$
years one can deduce
a limit on the grand unification scale $M_X\geq 10^{15}$~GeV [5].
Correspondingly,
supersymmetric GUT's are in a somewhat better shape than standard
GUT's although I would not call this evidence
for supersymmetry.
 
More in general,
it is quite difficult to tell where nonstandard
physics will first be seen. It can either appear in the form of new
unexpected particles or in the form of unexpected behavior of one
of the known particles.
Up to now, all experiments in these directions have
turned out to be rather frustrating because they are null experiments
looking for tiny deviations of the standard model (e.g.\ proton
decay, neutrino masses etc.). From a theoretical point of view
high energy experiments are generally superior to low energies
because the effect of new physics typically grows like a power
of $\frac{E}{\Lambda}$, $E$ being the energy involved. This is
a strong argument for
all ultrahigh energy colliders, in addition to top quark and
Higgs physics.
 
Within the standard model,
the Higgs boson is the most speculative particle.
In my opinion, it is possible that it does not exist at all,
so that the standard model would have to be replaced by another
theory with different high energy phenomenology.
In any case, once the Higgs particle is found, large
deviation from the standard model
might quickly show up (in the form of
several Higgs fields or else). As discussed before these could be
studied in high energy collisions.
 
Together with the top quark and the vector bosons
the Higgs field is the heaviest of the known particles (with
masses of order 100~GeV). Within the standard model all these particles
are pointlike. In my opinion it is difficult to imagine how
particles as heavy as a large nucleus can be pointlike to the
same extend than the light fermions (electron and up and down
quark). Although I cannot prove it,
I am expecting deviations from pointlike behavior to be
seen at the level of TeV energy experiments.
This view is supported by the fact that $m_t$ and $m_H$ are of the
same order of magnitude as the new physics scale $\Lambda\sim O$~(1~TeV)
mentioned above. If new physics exists at scales of order 1~TeV, this will
be seen first by the experimental analysis of top quark and Higgs boson
properties.
 
In summary, all theoretical developments beyond the 
standard model are not
very definite. Experimenters
are well advised to keep their eyes open
for a wide variety of possibilities. High energy precision experiments
at LHC and NLC seem to be most promising.\\
 
{\bf 9. Neutrino masses and mixings}
 
\begin{figure}
\begin{center}
\epsfig{file=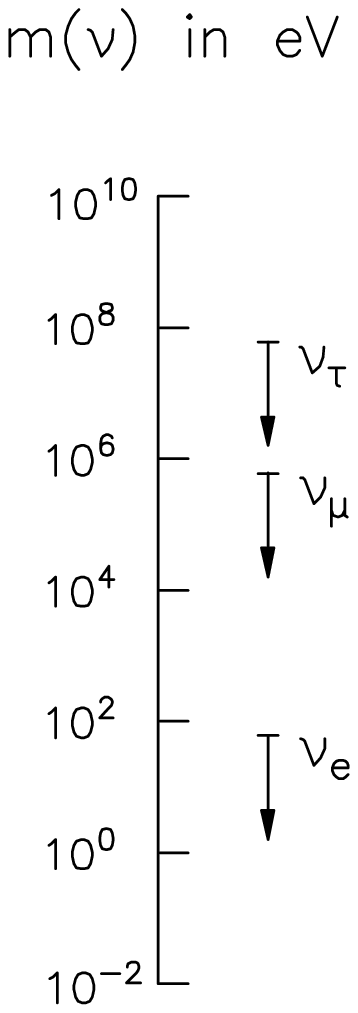,height=5cm}
\vskip 0.5cm
\caption{}
\label{fig12}
\end{center}
\end{figure}

Experimentally, all neutrino masses are tiny, if not zero. The
present experimental upper bounds for the 3 neutrino species are given
in fig. \ref{fig12}.
 
In the standard model
all neutrinos are massless. However, there is no fundamental
theoretical reason for that and, in fact, the standard model
can easily be extended
to include small neutrino masses. These masses can be either of
Dirac or of Majorana type, because the neutrino is electrically neutral.
The main question from the theoretical point of view is to understand the
small masses of neutrinos as compared to other fermions. As yet
no real
answer to this question has been given. Therefore, neutrino masses
may have any value, from extremely tiny to the upper limits given in
fig. \ref{fig12}. The ignorance about $m_\nu$ may be parametrized in
the form of $m_\nu=\frac{m^2}{M}$ where $m$ is
a typical fermion mass of order GeV and M a large unknown
scale.
 
Very small neutrino masses cannot be determined directly but show
up in the form of oscillations between the various
neutrino species. Thus a large number of experiments
searching for $\nu_e-\nu_\mu$ oscillations has been done,
sensitive to neutrino mass differences down to the eV range. However,
the present experiments provide no evidence for neutrino
masses. In addition, astrophysical observations and
cosmological considerations have led to no conclusion about
neutrino masses which I would take seriously.
 
There is only one statement about neutrino masses which I consider
most probably true: namely, if the neutrinos have
masses, the $\tau$-neutrino will be the neutrino with the
largest mass. Any experiment which aims at $m(\nu_\tau)$ should have
priority to other neutrino-experiments. Thus NOMAD and
CHORUS at CERN and
P803 at Fermilab are important and reasonable projects [6].\\  
 
{\bf 10. Conclusions}
 
Since I have a clearcut message I will make my conclusions very
short. We have seen that there are several interesting items in
future elementary particle physics.
By and large, one can be content with the direction
high energy physics takes.
Many interesting experiments (HERA, LEP2, LHC, P803, ...)
are under way and the
SSC cancellation is not unwise
because most of the SSC physics will be
covered by LHC, at a much lower price.
However, I have objections at certain specific points.
For example, one should not build
two $b$-factories where one machine would be enough.
Furthermore, I do not understand the widely spread attitude
of hesitation towards the really new large projects, like
the NLC.
 
Among all the items discussed, I would classify only one
as being {\em extremely} urgent and 
important. This is the question
about the heavy sector of the standard model,
i.e.\ the top quark and the Higgs
boson. I am quite sure that in the behavior of these particles physics
beyond the standard model
will show up and that one can get insight
into more fundamental laws of nature. Therefore every effort
should be made to study top quark and Higgs boson precisely,
by means of LHC and NLC.
This is certainly a very expensive program.
However, I see no reason to wait and to do
less important but cheaper experiments.
I may not be fully objective here
and admit that I am impatient. I am impatient to learn everything
about the smallest distances, and I passionately believe that the real
problems are still in front of us.


\begin{thebibliography}{}
\bibitem{Treille} see e.g. D.~Treille, 
        The LEP~200 program,CERN-PPE-93-54-REV(1993).
\bibitem{Iwata} see e.g. S. Iwata, KEK-93-157 (1993);\\
       V. Luth, Talk presented at the 1994 meeting of the APS,
       Albuquerque 1994.
\bibitem{Jarlskog} G. Jarlskog, D. Rein (eds), Proc.\ of the LHC
       Workshop, Aachen 1990.
\bibitem{Zerwas} P.M. Zerwas (ed), Proc. of the Workshops on
       $e^+e^-$ Collisions at 500~GeV, Hamburg 1993.
\bibitem{Revol} for a recent summary of future proton decay 
       experiments see J.P. Revol, Talk presented at the first
       International Conference on Phenomenology of Unification,
       Rome 1994.
\bibitem{Chorus} CHORUS collaboration, N. Armenise et al.,
      CERN-SPSC/90-42 (1990);\\
      NOMAD collaboration, P. Astier et al., CERN-SPSC/91-21 (1991);\\ 
      FNAL proposal P803, K. Kodama et al. (1991) .
\end{thebibliography}
\end{document}